\documentclass[aps,prl,reprint,superscriptaddress]{revtex4-1}
\usepackage{graphicx}
\usepackage{amsfonts}
\usepackage{color,amsmath}
\usepackage[normalem]{ulem}
\usepackage[hypertexnames=false]{hyperref}
\usepackage{lipsum}

\pdfpageattr {/Group << /S /Transparency /I true /CS /DeviceRGB>>}
\newcommand{\Vt}{V_{\rm{top}}}

\newcommand{\Vtop}{V_{\rm{top}}}
\newcommand{\Vqpc}{V_{\rm{qpc}}}
\newcommand{\Bp}{B_{\parallel}}
\newcommand{\Bt}{B_{\rm t}}
\newcommand{\Bc}{B_{\rm c}}
\newcommand{\Bo}{B_{\rm{\perp}}}
\newcommand{\Phio}{\Phi_{0}}
\newcommand{\Ez}{E_{\rm Z}}
\newcommand{\Et}{E_{\rm T}}

\newcommand{\Wso}{W_{\rm{S1}}}
\newcommand{\Wst}{W_{\rm{S2}}}
\newcommand{\xis}{\xi_{\rm{S}}}
\newcommand{\Vsd}{V_{\rm{sd}}}

\newcommand{\dGo}{G''(\Vsd =0)}
\newcommand{\dG}{G''(\Vsd )}

\newcommand{\It}{I_{\rm{3 \omega}}}
\newcommand{\Ito}{I_{\rm{3 \omega}}(\Vsd)}
\newcommand{\Itoo}{I_{\rm{3 \omega}}(\Vsd =0)}

\begin{document}

\title{Evidence of topological superconductivity in planar Josephson junctions}

\author{Antonio~Fornieri}
\altaffiliation{These authors contributed equally to this work.}
\affiliation{Center for Quantum Devices and Station Q Copenhagen, Niels Bohr Institute, University of Copenhagen, Universitetsparken 5, 2100 Copenhagen, Denmark}

\author{Alexander~M.~Whiticar}
\altaffiliation{These authors contributed equally to this work.}
\affiliation{Center for Quantum Devices and Station Q Copenhagen, Niels Bohr Institute, University of Copenhagen, Universitetsparken 5, 2100 Copenhagen, Denmark}

\author{F.~Setiawan}
\affiliation{James Franck Institute, The University of Chicago, Chicago, IL 60637, USA}

\author{El\'ias~Portol\'es Mar\'in}
\affiliation{Center for Quantum Devices and Station Q Copenhagen, Niels Bohr Institute, University of Copenhagen, Universitetsparken 5, 2100 Copenhagen, Denmark}

\author{Asbj\o rn~C.~C.~Drachmann}
\affiliation{Center for Quantum Devices and Station Q Copenhagen, Niels Bohr Institute, University of Copenhagen, Universitetsparken 5, 2100 Copenhagen, Denmark}


\author{Anna Keselman}
\affiliation{Station Q, Microsoft Research, Santa Barbara, California 93106-6105, USA}

\author{Sergei~Gronin}
\affiliation{Department of Physics and Astronomy and Station Q Purdue, Purdue University, West Lafayette, Indiana 47907 USA}
\affiliation{Birck Nanotechnology Center, Purdue University, West Lafayette, Indiana 47907 USA}

\author{Candice~Thomas}
\affiliation{Department of Physics and Astronomy and Station Q Purdue, Purdue University, West Lafayette, Indiana 47907 USA}
\affiliation{Birck Nanotechnology Center, Purdue University, West Lafayette, Indiana 47907 USA}

\author{Tian~Wang}
\affiliation{Department of Physics and Astronomy and Station Q Purdue, Purdue University, West Lafayette, Indiana 47907 USA}
\affiliation{Birck Nanotechnology Center, Purdue University, West Lafayette, Indiana 47907 USA}

\author{Ray~Kallaher}
\affiliation{Birck Nanotechnology Center, Purdue University, West Lafayette, Indiana 47907 USA}
\affiliation{Microsoft Quantum at Station Q Purdue, Purdue University, West Lafayette, Indiana 47907, USA}

\author{Geoffrey~C.~Gardner}
\affiliation{Birck Nanotechnology Center, Purdue University, West Lafayette, Indiana 47907 USA}
\affiliation{Microsoft Quantum at Station Q Purdue, Purdue University, West Lafayette, Indiana 47907, USA}

\author{Erez~Berg}
\affiliation{James Franck Institute, The University of Chicago, Chicago, IL 60637, USA}
\affiliation{Department of Condensed Matter Physics, Weizmann Institute of Science, Rehovot 7610001, Israel}

\author{Michael~J.~Manfra}
\affiliation{Department of Physics and Astronomy and Station Q Purdue, Purdue University, West Lafayette, Indiana 47907 USA}
\affiliation{Birck Nanotechnology Center, Purdue University, West Lafayette, Indiana 47907 USA}
\affiliation{School of Materials Engineering, Purdue University, West Lafayette, Indiana 47907 USA}
\affiliation{School of Electrical and Computer Engineering, Purdue University, West Lafayette, Indiana 47907 USA}

\author{Ady~Stern}
\affiliation{Department of Condensed Matter Physics, Weizmann Institute of Science, Rehovot 7610001, Israel}

\author{Charles~M.~Marcus}
\email[email: ]{marcus@nbi.ku.dk}
\affiliation{Center for Quantum Devices and Station Q Copenhagen, Niels Bohr Institute, University of Copenhagen, Universitetsparken 5, 2100 Copenhagen, Denmark}

\author{Fabrizio~Nichele}
\email[email: ]{fni@ibm.zurich.com}
\altaffiliation{Present address: IBM Research - Zurich, Säumerstrasse 4, 8803 Rüschlikon, Switzerland.}
\affiliation{Center for Quantum Devices and Station Q Copenhagen, Niels Bohr Institute, University of Copenhagen, Universitetsparken 5, 2100 Copenhagen, Denmark}



\maketitle

\begin{figure}[t]
\includegraphics[width=\columnwidth]{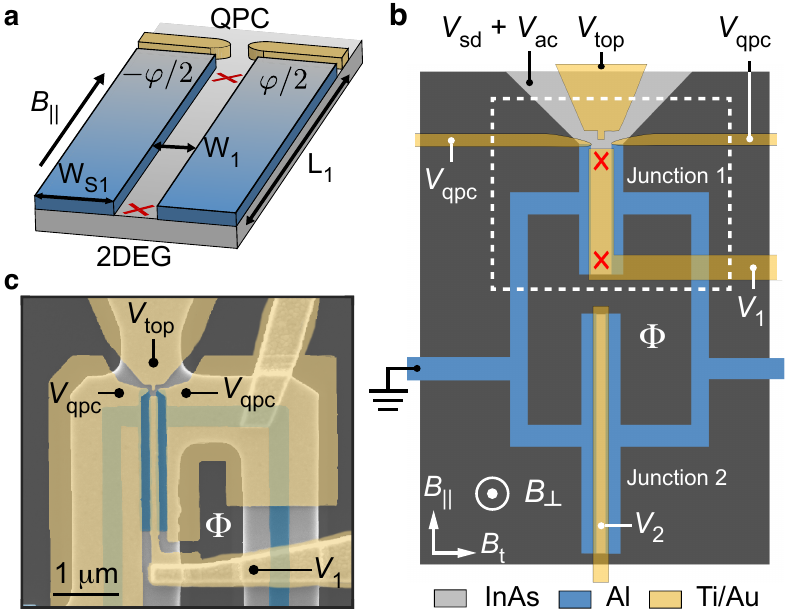}
\caption{\textbf{Topological Josephson junction.} \textbf{a}, Schematic of a planar JJ formed by two epitaxial superconducting layers (represented in blue) covering a 2DEG with strong spin-orbit interaction (grey). A one-dimensional channel, defined between the superconducting leads, can be tuned into the topological regime with Majorana modes (red crosses) at its ends by the parallel field $\Bp$, the 2DEG chemical potential $\mu$ and the phase difference between the superconductors $\varphi$. Majorana modes can be probed in tunneling spectroscopy using a QPC located at one end of the JJ. \textbf{b}, Schematic of the measured device (not to scale) consisting of a superconducting loop interrupted by two JJs (labeled 1 and 2) in parallel. The interferometer is formed by InAs 2DEG (light grey) and epitaxial Al (blue). Five Ti/Au gates (yellow) allow independent tuning of the chemical potential in JJ1 (gate voltage $V_1$), the chemical potential in JJ2 ($V_2$) and the transmission of a tunnel barrier at the top end of JJ1 ($\Vqpc$ and $\Vtop$). The applied AC and DC bias voltages are also indicated, together with the direction of magnetic field parallel ($\Bp$) and transverse  ($B_{\rm t}$) to the JJ, and the magnetic flux $\Phi$. \textbf{c}, False color scanning electron micrograph of the top part of a typical device, as in the dashed box shown in panel \textbf{b}. The colors are the same as those used in panel \textbf{b}.}
\label{fig1}
\end{figure}

\textbf{\boldmath Majorana zero modes are quasiparticle states localized at the boundaries of topological superconductors that are expected to be ideal building blocks for fault-tolerant quantum computing~\cite{Kitaev2003,Nayak2008}. Several observations of zero-bias conductance peaks measured in tunneling spectroscopy above a critical magnetic field have been reported as experimental indications of Majorana zero modes in superconductor/semiconductor nanowires~\cite{Mourik2012,Das2012,Deng2016,Suominen2017,Nichele2017,Zhang2018}. 
On the other hand, two dimensional systems offer the alternative approach to confine Majorana channels within planar Josephson junctions, in which the phase difference $\varphi$ between the superconducting leads represents an additional tuning knob predicted to drive the system into the topological phase at lower magnetic fields~\cite{Hell2017,Pientka2017}. Here, we report the observation of phase-dependent zero-bias conductance peaks measured by tunneling spectroscopy at the end of Josephson junctions realized on a InAs/Al heterostructure. 
Biasing the junction to $\varphi \sim \pi$ significantly reduces the critical field at which the zero-bias peak appears, with respect to $\varphi=0$. The phase and magnetic field dependence of the zero-energy states is consistent with a model of Majorana zero modes in finite-size Josephson junctions.
Besides providing experimental evidence of phase-tuned topological superconductivity, our devices are compatible with superconducting quantum electrodynamics architectures~\cite{Casparis2018} and scalable to complex geometries needed for topological quantum computing~\cite{Hell2017,Hell2017b}.}

The Josephson junctions (JJs) studied in this work were fabricated from a planar heterostructure comprising of a thin Al layer epitaxially covering a high mobility InAs two-dimensional electron gas (2DEG) \cite{Shabani2015}. As a consequence of the highly transparent superconductor/semiconductor interface~\cite{Krogstrup2015}, a hard superconducting gap is induced in the InAs layer \cite{Kjaergaard2016,Kjaergaard2017}. Selectively removing an Al stripe of width $W_1$ and length $L_1$ defines a normal InAs region, laterally contacted by superconducting leads, as shown in Fig~\ref{fig1}a. Superconducting gaps $\Delta\; \mathrm{exp(\pm i \varphi /2)}$ opening below the Al planes on the right- and left-hand side~\cite{Blonder1982,Klapwijk2004}, respectively, confine low energy quasiparticles within the normal InAs channel.
Owing to the strong spin-orbit interaction in InAs~\cite{Shabani2015}, together with the lateral confinement, the JJ of Fig~\ref{fig1}a is predicted to undergo a topological transition at high magnetic field $\Bp$ parallel to the junction~\cite{Hell2017}, with Majorana modes isolated from the continuum forming at the end points (crosses in Fig~\ref{fig1}a), similarly to conventional nanowires \cite{Lutchyn2010,Oreg2010}. Most strikingly, phase control offers an additional tuning parameter to enter the topological regime not explored so far. Biasing the JJ to $\varphi=\pi$ was predicted to significantly reduce the critical magnetic field of the topological transition, and to enlarge its phase boundaries in chemical potential~\cite{Pientka2017}. 

Here we investigate planar JJs as that of Fig~\ref{fig1}a as a function of $\Bp$, chemical potential $\mu$ and phase difference $\varphi$. Phase biasing is obtained by embedding the JJ in a direct-current superconducting quantum interference device (DC SQUID) threaded by a magnetic flux~\cite{Tinkham}. A robust zero-bias peak (ZBP) exhibiting strong dependence on $\varphi$ is measured via tunneling spectroscopy using a laterally coupled quantum point contact (QPC), as schematically shown in Fig~\ref{fig1}a. The ZBP behavior is consistent with a Majorana mode in a finite-size topological JJ (see Extended Data Fig.~\ref{figEDF1}).

Figure~\ref{fig1}b shows a schematic of our device, which consists of a three-terminal asymmetric SQUID with two JJs, labeled 1 and 2, and a tunneling probe coupling to a normal lead on the top end of JJ1. Figure~\ref{fig1}c shows an electron micrograph in the surrounding of JJ1. The junctions are characterized by Josephson critical currents $I_{\rm c,2}> I_{\rm c,1}$, such that the phase difference  $\varphi$  across JJ1 can be tuned from 0 to $\sim \pi$ by threading the SQUID loop with a magnetic flux $\Phi$ (generated by the out-of-plane field $\Bo$) varying from 0 to $\Phi_0/2$, where $\Phi_0=h/2e$ is the superconducting flux quantum ($e$ is the electron charge and $h$ the Planck constant). The SQUID is laterally connected to two superconducting leads that serve as ground and allow measuring the Josephson critical current of the interferometer (see Extended Data Figs.~\ref{figEDF10} and~\ref{figEDF11}). The SQUID loop is obtained by a combination of deep wet etching on the semiconductor heterostructure and selective wet etching of the top Al layer. A $\mathrm{HfO_2}$ dielectric layer is deposited over the entire sample for gate isolation, followed by lift-off of the Ti/Au gate structures. Top gates $V_1$ and $V_2$ control the chemical potential in JJ1 and JJ2, respectively. Split gates deposited at the top end of JJ1 form a QPC. In the tunneling regime, the QPC serves as spectroscopic probe revealing the local density of states of JJ1. The uppermost gate extends between the QPC gates and helps defining a sharp tunnel barrier when operated at a voltage $\Vtop \sim 0$. To ensure a hard superconducting gap in high parallel fields, the QPC gates additionally confine the 2DEG beneath the narrow Al leads~\cite{Suominen2017,Nichele2017} (see Fig~\ref{fig1}c). We present data for a device with $W_{\rm 1}=80$, $W_{\rm 2}=40$ nm, $L_{\rm 1}=1.6~\rm{\mu m}$ and $L_{\rm 2}=5~\rm{\mu m}$. The width of the superconducting leads is $\Wso=\Wst=160$ nm for both JJs, and SQUID loop area $\sim 8~\rm{\mu m}^2$. Data was reproduced for two additional devices with $W_1=80$ and $120$ nm respectively and presented in Extended Data Figs.~\ref{figEDF7}-\ref{figEDF9}.
Differential conductance $G$ was measured in a four-terminal configuration by standard AC lock-in techniques in a dilution refrigerator with an electron base temperature of about 40~mK. 

\begin{figure*}[t]
\includegraphics[width=\textwidth]{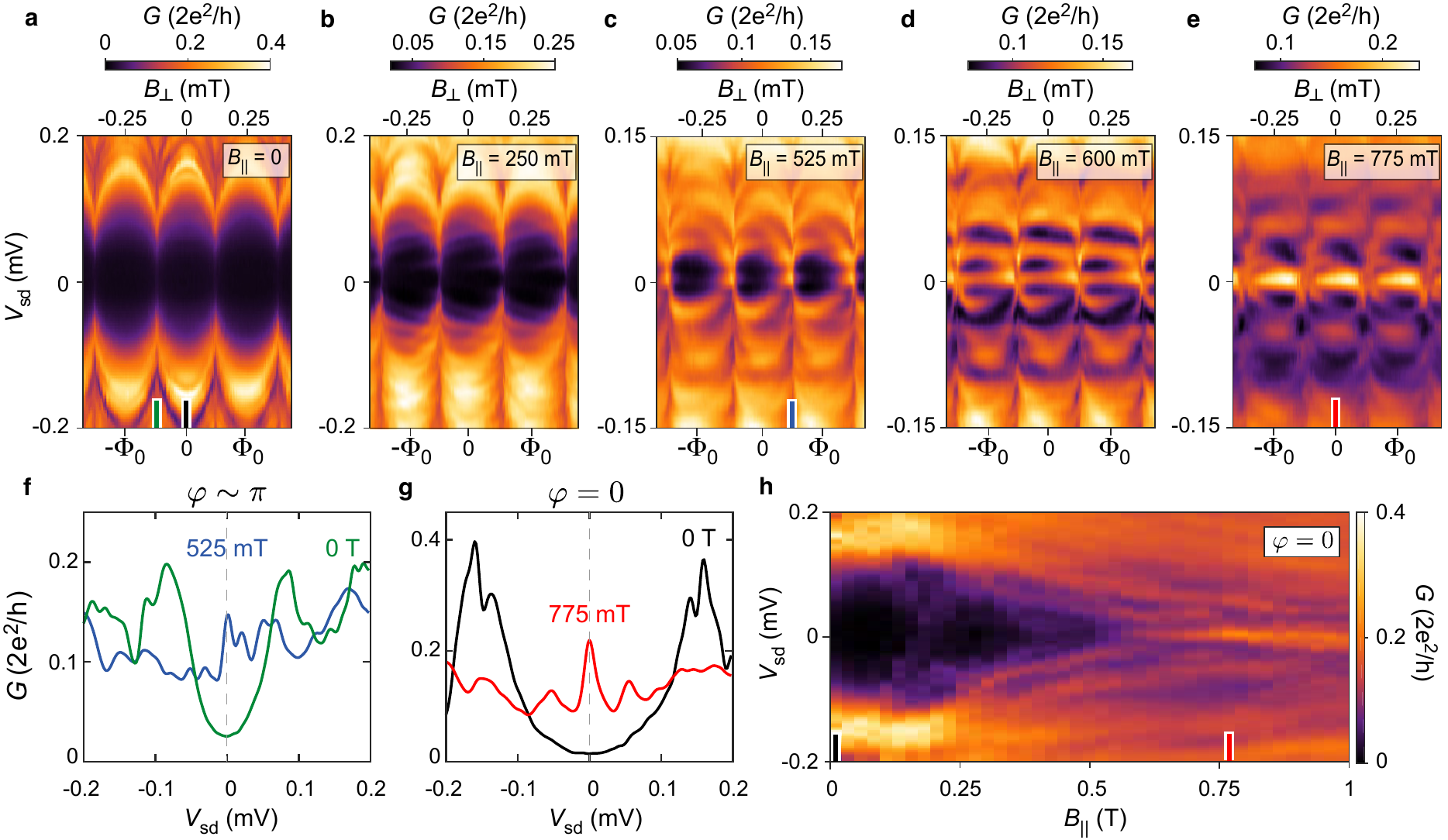}
\caption{\textbf{Evolution of the zero-bias peak in parallel field.} \textbf{a-e},  Differential conductance $G$ as a function of magnetic flux, $\Phi$, piercing the SQUID loop and source-drain bias, $\Vsd$, measured at different values of magnetic field, $\Bp$, parallel to the junction. The flux is generated by the out-of-plane field $\Bo$. The values of $\Bo$ have been shifted to remove offsets. \textbf{f}, Line-cuts of $G$ versus $\Vsd$ at phase bias $\varphi \sim \pi$ for $\Bp=0$ (green line) and $\Bp=525$ mT (blue line), as indicated by the ticks in panels \textbf{a} and \textbf{c}. \textbf{g}, Line-cuts of $G$ versus $\Vsd$ at $\varphi=0$ for $\Bp=0$ (black line) and $\Bp=775$ mT (red line), as indicated by the ticks in panels \textbf{a} and \textbf{e}. \textbf{h}, $G$ as a function of $\Vsd$ and $\Bp$ at $\varphi=0$. The plot was reconstructed from line-cuts as the ones shown in panel \textbf{g}. The measurements were taken at the top gate voltage $V_1=V_1^{\star}=-118.5$ mV.}
\label{fig2}
\end{figure*}

Figure~\ref{fig2}a shows $G$ as a function of the bias voltage $\Vsd$ and $\Phi$ at $\Bp=0$. The superconducting gap $\Delta (\Phi = 0) \simeq 150~\mathrm{\mu}$eV periodically oscillates as a function of $\Phi$ and is reduced by approximately 50 \% at $\Phi =(2n+1)\Phio/2$, where $n$ is an integer. This behavior indicates phase-coherent transport through JJ1 generated by Andreev reflection processes~\cite{Andreev1964,Kulik1970} at the interfaces between the bare 2DEG and the proximitized leads. The flux modulation of the whole continuum of states outside the gap is expected for JJs with narrow superconducting leads ($\Wso \ll \xis$, where $\xis=\hbar v_{\rm F}/\pi \Delta \sim 1.5~\mathrm{\mu}$m is the superconducting coherence length and $v_{\mathrm F}$ is the Fermi velocity in the semiconductor), while the non-complete closure of the gap at $\Phi =(2n+1)\Phio/2$ is associated to the finite length $L_1$ of the junction (see Methods and Extended Data Fig.~\ref{figEDF1} for further details). 

As $\Bp$ is increased, discrete Andreev bound states (ABSs) enter the gap and move towards zero energy, as shown in Fig.~\ref{fig2}b for $\Bp=250$ mT. These states have an asymmetric flux dependence, as expected for JJs with strong spin-orbit interaction in the presence of a Zeeman field~\cite{Yokoyama2013,VanWoerkom2017}. 

For $\Bp=525$ mT a ZBP in conductance appears at $\Phi=(2n+1)\Phi_0/2$ (corresponding to $\varphi\sim \pi$, see Fig.~\ref{fig2}f), while it vanishes when $\Phi$ is set to $2n\Phi_0$, i.e. when $\varphi=0$, as shown in Fig.~\ref{fig2}c. As $\Bp$ reaches 600 mT the ZBP instead extends over the whole $\varphi$ range, except at $\varphi \sim \pi$ where it moves to finite bias (Fig.~\ref{fig2}d). The state sticks at zero energy as the field is increased around $\varphi= 0$, while it splits and moves to higher energies at $\varphi\sim \pi$. This behavior is observable for instance in Fig.~\ref{fig2}e, in which the conductance shows a minimum at $\varphi \sim \pi$ while a bright zero-energy state extends over the rest of the phase range. The conductance line-cuts for $\varphi \sim \pi$ and $\varphi = 0$ at different values of $\Bp$ are shown in Figs.~\ref{fig2}f and~\ref{fig2}g, respectively. The full field evolution of the zero-energy state for $\varphi=0$ is shown in Fig.~\ref{fig2}h. Above $\Bp=1$ T the induced gap softens and the phase dependence of sub-gap states gradually disappears as the JJs of the SQUID reach the resistive state.

The observed behavior of the ZBP in field and phase is in good qualitative agreement with the calculated spectrum of a topological JJ with realistic parameters, as shown in Extended Data Fig.~\ref{figEDF1}. The calculated gapped zero-energy state is characterized by a Majorana wave function localized at the edges of the JJ (see Extended Data Fig.~\ref{figEDF1}h). The observed splitting at $\varphi \sim \pi$ is reproduced by the simulations and can be understood in terms of hybridization of the Majorana modes, since at this value of $\varphi$ the induced gap is minimized and, as a result, the coherence length is maximized.

\begin{figure*}
\includegraphics[width=\textwidth]{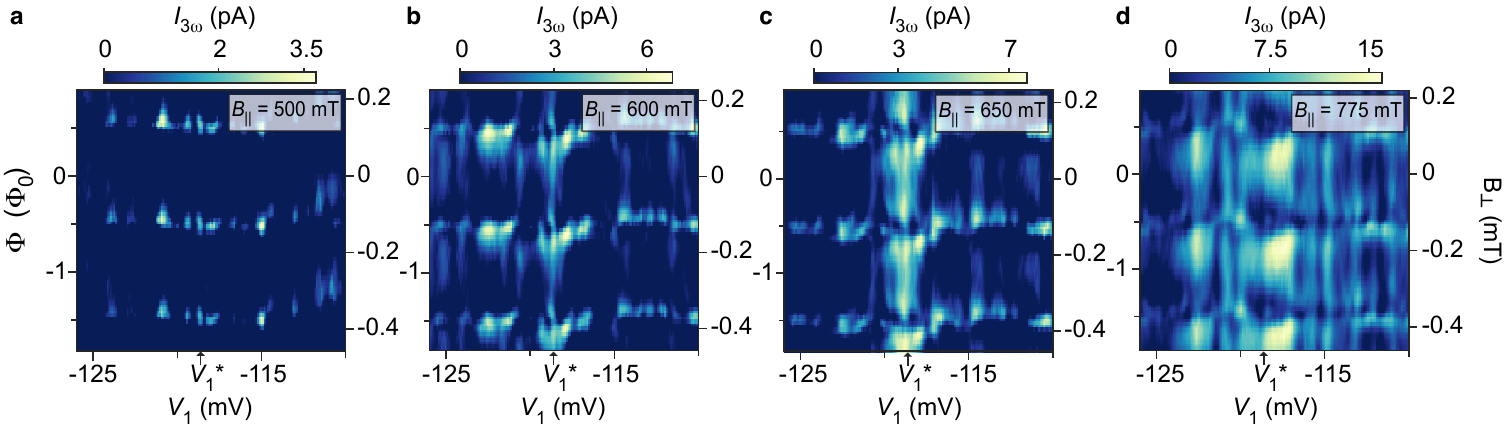}
\caption{\textbf{Stability of the zero-bias peak.} \textbf{a-d}, Third harmonic of the current $\Itoo$ measured by the lock-in amplifier at zero bias as a function of top gate voltage, $V_1$, and magnetic flux, $\Phi$, for different values of magnetic field, $\Bp$, parallel to the junction. The flux is generated by the out-of-plane field $\Bo$. The values of $\Bo$ have been shifted to remove offsets. As shown in the Methods and Extended Data Fig.~\ref{figEDF6}, $\Itoo \propto -\dGo$, where $\dGo=(\partial^2 G/\partial V^2)|_{\Vsd=0}$ is the second derivative of the conductance at zero bias. A positive value of $\Itoo$ corresponds to a ZBP in conductance as a function of $\Vsd$. 
}
\label{fig3}

\end{figure*}
One of the most interesting features predicted for a perfectly transparent JJ is the significant expansion of the topological phase in magnetic field and chemical potential at $\varphi=\pi$~\cite{Pientka2017}. We therefore investigated the stability of the  ZBP, starting from its dependence on the gate voltage $V_1$, which controls the chemical potential in JJ1. In order to explore efficiently our 4-dimensional parameter space, we recorded the third harmonic $\Ito$ of the current measured by the lock-in amplifier.  As shown in the Methods and in Extended Data Fig.~\ref{figEDF6}, $\Ito \propto -\dG=-(\partial^2 G/\partial V^2)|_{\Vsd}$. A ZBP in conductance is therefore identified by a positive value of $\Itoo$, i.e., by a negative value of $\dGo$, which indicates a negative curvature around $\Vsd = 0$. 

Figure~\ref{fig3} displays $\Itoo$ as a function of $\Phi$ and $V_1$ for different values of $\Bp$. At $\Bp=500$ mT horizontal stripes showing positive values of $\Itoo$ are visible at $\varphi \simeq (2n+1)\pi$. Increasing the field causes the region of negative curvature to expand around the voltage $V_1^{\star}=-118.5$ mV by $\sim 2$ mV and in phase extending to $2n\pi$. For $\Bp=650$ mT the ZBP region expands further around $V_1^{\star}$, while a maximum develops at $\varphi\simeq(2n+1)\pi$, indicating that the ZBP has split to finite energy. The ZBP region covers a maximum range of 10 mV at 775 mT and remains extended in phase for $\varphi\neq (2n+1)\pi$.

The finite range of $V_1$ over which the ZBP is stable is explained by the narrow width $\Wso\ll \xis$ of the superconducting leads, which effectively decrease the ratio between Andreev and normal reflection probabilities, thus reducing the size of the topological phase as a function of $\mu$ (see Extended Data Fig.~\ref{figEDF1}a). Although this geometry causes a deviation from the predicted behavior of a topological JJ, in our devices the finite width is necessary to guarantee a well-defined induced gap up to 1 T (see Methods for further details).

\begin{figure}
\includegraphics[width=\columnwidth]{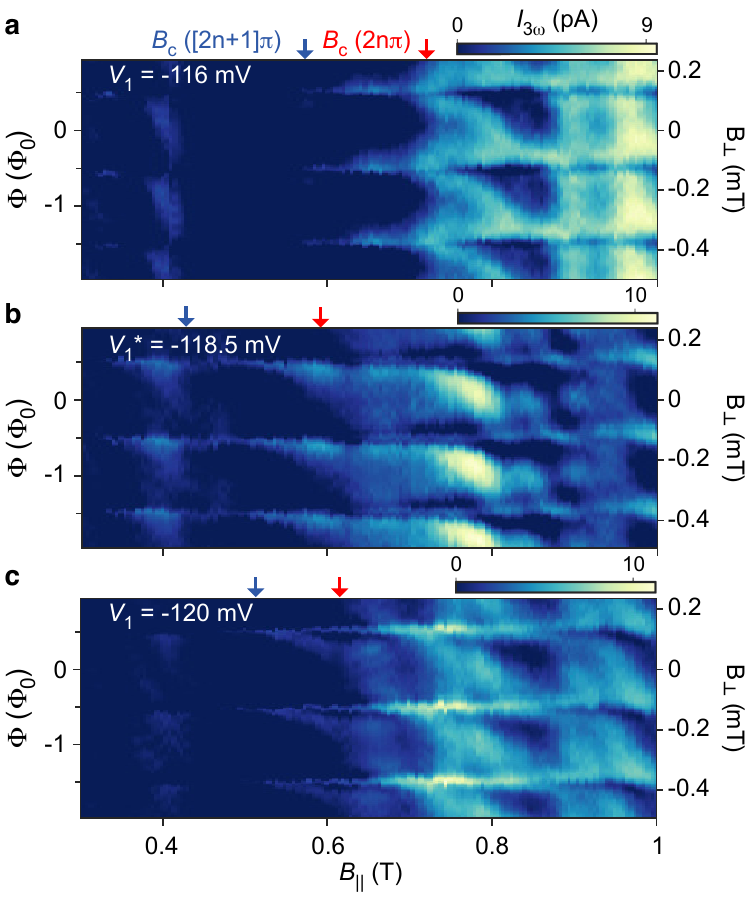}
\caption{\textbf{Phase dependence of the critical field.} \textbf{a-c}, Third harmonic of the current $\Itoo$ measured by the lock-in amplifier at zero bias as a function of magnetic field, $\Bp$, applied parallel to the junction and magnetic flux, $\Phi$, for different values of top gate voltage $V_1$. The flux is generated by the out-of-plane field $\Bo$. The values of $\Bo$ have been shifted to remove offsets. A positive value of $\Itoo$ corresponds to a ZBP in conductance as a function of $\Vsd$. 
}
\label{fig4}
\end{figure}

The complementary study of the ZBP stability in $\Phi$ and $\Bp$ for different values of $V_1$ is shown in Fig.~\ref{fig4}. At $V_1=-116$ mV (Fig.~\ref{fig4}a) extended regions of positive $\Itoo$ indicating a stable ZBP appear above an oscillating critical field $\Bc(\varphi)$, which reaches a minimum value $\Bc([2n+1]\pi) = 570$ mT, as indicated by the blue arrow. On the other hand, the vertical stripe visible at $\sim 0.4$ T is due to ABSs crossing zero energy without sticking. Similarly to what was observed above as a function of the chemical potential, the negative curvature region significantly expands in terms of $\Bp$ range for $V_1=V_1^{\star}$, where $\Bc([2n+1]\pi)  = 435$ mT (Fig.~\ref{fig4}b). At more negative $V_1$ the ZBP regime contracts again ($\Bc([2n+1]\pi) = 480$ mT, Fig~\ref{fig4}c), consistent with the stability maps shown in Fig.~\ref{fig3}. 

This behavior is in qualitative agreement with the topological phase diagrams calculated for our system (see Extended Data Fig.~\ref{figEDF1}a,b). Furthermore, variations of $V_1$ evidently modify the amplitude of $\Bc$ modulations, which can reach a maximum visibility $\Bc(2n \pi)-\Bc([2n+1] \pi)/\Bc([2n+1] \pi)\sim 37 \%$. We note that the strong asymmetry in the phase dependence of the ZBP region observed at different values of $V_1$ is not fully captured by the numerical model and might be due to unintended asymmetries in the geometry of the etched superconducting leads. 
Lastly, we note that the ZBP is robust over a range of $\sim 70$ mV in $\Vqpc$ and in $\Vt$, which modify the above-gap conductance by approximately one order of magnitude, as shown in Extended Data Fig.~\ref{figEDF3}. 

As another interesting test of the topological nature of the observed ZBPs, we performed spectroscopy as a function of the magnetic field $\Bt$ applied in the plane of the 2DEG but orthogonal to the one dimensional channel defined by JJ1 (see Fig.~\ref{fig1}b). In this field orientation we do not observe any discrete state sticking at zero energy before the suppression of the induced gap, which occurs at $\Bt\sim 360$ mT (see Extended Data Fig.~\ref{figEDF2}). 

Finally, it is worth noting that a first-order topological transition is expected for a planar JJ in presence of strong parallel fields. This transition should manifest itself with a minimum of the Josephson critical current~\cite{Hart2016} when the Zeeman energy reaches a value comparable to the Thouless energy $\Et$~\cite{Pientka2017}. In our case, however, this limit cannot be reached since $\Et\sim 2.8$ meV is one order of magnitude larger than $\Delta$. Experimentally, we observed periodic revivals of the Josephson current flowing through JJ1 as a function of $\Bp$, with a periodicity correlated to $\Wso$ (see Extended Data Fig.~\ref{figEDF11}). We ascribe this behavior to trivial orbital effects of the in-plane magnetic field in the proximitized 2DEG underneath the superconducting leads~\cite{Pientka2017}.

In summary, we have investigated phase-dependent ZBPs in tunneling conductance measured at the edge of a JJ patterned in a two-dimensional InAs/Al heterostructure. The critical field at which the ZBP appears depends on the phase bias and is minimal at $\varphi \sim \pi$, as expected for a topological JJ. We studied the ZBP stability as a function of field $\Bp$, phase $\varphi$ and chemical potential $\mu$, obtaining results qualitatively consistent with the topological phase diagram of a finite-size junction. Future material and design improvements might allow the investigation of JJs with $\Wso \gg \xis$, where the influence of $\mu$ is expected to be suppressed~\cite{Pientka2017}. Together with our top-down fabrication approach, the phase tuning of topological channels would extremely simplify the realization of sophisticated network geometries required to implement topologically protected quantum devices.

\textbf{Acknowledgments.} This work was supported by Microsoft Corporation, the Danish National Research Foundation, the Villum Foundation, ERC Project MUNATOP, CRC-183 and the Israeli Science Foundation. We thank Eoin O'Farrell, Michael Hell, Karsten Flensberg and Joshua Folk for useful discussions.

\textbf{Author contributions.} S.G., C.T., T.W., R.K., G.C.G. and M.J.M. developed and grew the InAs/Al heterostructure. A.F., A.M.W. and A.C.C.D. fabricated the devices. A.F., A.M.W. and A.C.C.D. performed the measurements with input from C.M.M. and F.N. Data analysis was done by A.F., A.M.W. and E.P.M. F.S., A.K., E.B. and A.S. developed the theoretical model and carried out the simulations. C.M.M. and F.N. conceived the experiment. All authors contributed to interpreting the data. The manuscript was written by A.F., A.M.W. and F.N. with suggestions from all other authors.

\section*{Methods}
\setcounter{equation}{0}
\setcounter{figure}{0}
\setcounter{table}{0}
\makeatletter
\renewcommand{\figurename}{EXTENDED DATA FIG.}
\textbf{Wafer structure.} The wafer structure used for this work was grown on an insulating InP substrate by molecular beam epitaxy. From bottom to top, it is comprised of a 100-nm-thick $\rm{In_{0.52}Al_{0.48}As}$ matched buffer, a 1-$\mathrm{\mu m}$-thick step-graded buffer realized with alloy steps from $\rm{In_{0.52}Al_{0.48}As}$ to $\rm{In_{0.89}Al_{0.11}As}$ (20 steps, 50 nm/step), a $58~\rm{nm}$ $\rm{In_{0.82}Al_{0.18}As}$ layer, a $4~\rm{nm}$ $\rm{In_{0.75}Ga_{0.25}As}$ bottom barrier, a $7~\rm{nm}$ InAs quantum well, a $10~\rm{nm}$ $\rm{In_{0.75}Ga_{0.25}As}$ top barrier, two monolayers of GaAs and a $7~\rm{nm}$ film of epitaxial Al. The top Al layer has been grown in the same molecular beam epitaxy chamber used for the rest of the growth, without breaking the vacuum. This result in semiconductor/superconductor interfaces characterized by almost unitary transparency \cite{Kjaergaard2017}. The two monolayers of GaAs are introduced to help passivate the wafer surface where the Al film is removed, and to make the sample more compatible with our Al etchant, which does not attack GaAs. The 2DEG is expected to mainly reside in the InAs quantum well, with the upper tail of the wavefunction extending to the Al film \cite{Shabani2015}.

Characterization performed in a Hall bar geometry where the Al was removed revealed an electron mobility peak $\mu=43,000~\rm{cm^2V^{-1}s^{-1}}$ for an electron density $n=8\times10^{11}~\rm{cm^{-2}}$, corresponding to an electron mean free path of $\sim 600$ nm. Characterization of a large area Al film revealed a critical magnetic field of $2.5~\rm{T}$ when the field is applied in the plane of the 2DEG.

\textbf{Device fabrication.} Samples were fabricated with conventional electron beam lithography techniques. First, we isolated large mesa structures by locally removing the top Al layer (with Al etchant Transene D) and performing a deep III-V chemical wet etch (220:55:3:3 $\rm{H_2 O:C_6 H_8 O_7:H_3 PO_4:H_2 O_2}$). In a subsequent step, we patterned the Al SQUID by selectively removing the top Al layer with a wet etch (Al etchant Transene D) at a temperature of $(50^\circ\pm1)^\circ \, \mathrm{C}$ for $5~\rm{s}$.
We then deposit on the entire sample a $15~\rm{nm}$ thick layer of insulating $\rm{HfO_2}$ by atomic layer deposition at a temperature of $90^\circ \, \mathrm{C}$. The top gate electrodes are deposited in two successive steps. First we define the features requiring high accuracy and deposit $5~\rm{nm}$ of Ti and $25~\rm{nm}$ of Au by electron beam evaporation. In a successive step we define the gates bonding pads by evaporating $10~\rm{nm}$ of Ti and $250~\rm{nm}$ of Au. Ohmic contacts to the InAs are provided by the epitaxial Al layer, which is contacted directly by wedge bonding through the insulating $\rm{HfO_2}$. 

\textbf{Measurements.} Electrical measurements were performed in a dilution refrigerator with a base temperature of $15$ mK with conventional DC and lock-in techniques using low frequencies ($\nu <200~\rm{Hz}$) excitations. In order to measure the differential conductance $G=\mathrm{d} I/\mathrm{d}V$, an AC voltage bias of $V_{\rm ac}=3~\rm{\mu V}$, superimposed to a variable DC bias $V_{\rm sd}$, was applied to the top lead of the device (see Fig.~\ref{fig1}b), with one of the SQUID leads grounded via a low-impedance current-to-voltage converter. An AC voltage amplifier with an input impedance of $500~\rm{M\Omega}$ was used to measure the four terminal voltage across the device. Information about the second derivative of the conductance $\dG=\partial^2 G/\partial V^2 |_{\Vsd}$ was experimentally obtained by recording the third harmonic $\Ito$ of the current measured by the lock-in amplifier (model SR830, which allows to detect signals at harmonics of the reference frequency). Indeed, when a sinusoidal time-dependent excitation $V(t)=\Vsd+V_{\rm ac}\, \mathrm{sin}(\omega t)$ is applied to the device, the measured output current can be expanded in Taylor's series as:

\begin{align}
I(t) & \simeq I(\Vsd)+\frac{\partial I}{\partial V}\bigg |_{\Vsd} V_{\rm ac}\,\mathrm{sin}(\omega t)+\frac{1}{2}\frac{\partial^2 I}{\partial V^2}\bigg |_{\Vsd}\nonumber \\ 
&\times [V_{\rm ac}\,\mathrm{sin}(\omega t)]^2
 +\frac{1}{6}\frac{\partial^3 I}{\partial V^3}\bigg |_{\Vsd} [V_{\rm ac}\,\mathrm{sin}(\omega t)]^3.
\end{align}

Since $[\mathrm{sin}(\omega t)]^3=1/4[-\mathrm{sin}(3\omega t)+3\,\mathrm{sin}(\omega t)]$, we obtain:
\begin{align}
\Ito=-\frac{1}{24}\frac{\partial^3 I}{\partial V^3}\bigg |_{\Vsd}V_{\rm ac}^3\propto -\frac{\partial^2 G}{\partial V^2}\bigg |_{\Vsd}.
\end{align}
In order to increase the signal-to-noise ratio, the measurement of $\Ito$ was performed with an amplitude $V_{\rm ac}$ of the excitation greater than the temperature-limited full width at half maximum of a Lorentzian feature, i.e., $V_{\rm ac}\gtrsim 3.5\, k_{\rm B}T$, where $k_{\rm B}$ is the Boltzmann constant and $T\sim 40$ mK is the electron temperature in our devices. The comparison between $\Ito$ and $\dG$ is shown in Extended Data Fig.~\ref{figEDF6}.

The SQUID device was separately characterized. The SQUID differential resistance $R=dV/dI$ was obtained by applying an AC current bias $I_{\rm ac}<5$ nA, superimposed to a variable DC current bias $I_{\rm DC}$, to the superconducting leads of the interferometer, with the tunnneling probe closed and floating. The behavior of the Josephson critical current is shown in Extended Data Figs.~\ref{figEDF10} and~\ref{figEDF11}. 

We studied five devices characterized by different dimensions of JJ1. Devices 1 and 2 have $W_1=80$ nm and $\Wso=160$ nm, device 3 has $W_1=120$ nm and $\Wso=160$ nm, device 4 has $W_1=80$ nm and $\Wso=500$ nm, whereas device 5 has $W_1=80$ nm and $\Wso=1~\mathrm{\mu}$m. All the devices were designed with $L_1=1.6~\mathrm{\mu}$m, $W_2=40$ nm, $L_2=5~\mathrm{\mu}$m and  $\Wst=160$ nm. In device 3 spectroscopy was performed by means of a QPC coupling JJ1 to a wide Al plane, following the approach of Refs.~\citenum{Suominen2017,Nichele2017} (see Extended Data Fig.~\ref{figEDF9} for further details). Results consistent with those presented in the main text were obtained in devices 1, 2 and 3, while in devices 4 and 5 the induced superconducting gap collapsed at $\Bp \sim 200$ mT without showing any robust ZBP. The behavior of devices 4 and 5 is consistent with the softening of the induced gap in low parallel fields observed below wide superconducting leads~\cite{Kjaergaard2016}. 

\textbf{Theoretical model.} We model JJ1 of the measured device using the Hamiltonian~\cite{Pientka2017,Hell2017} written in the Nambu basis $(\psi_\uparrow,\psi_{\downarrow},\psi_{\downarrow}^\dagger,-\psi_{\uparrow}^\dagger)^T$ as
\begin{align}\label{eq:HSM}
H &= \left(-\frac{\hbar^2(\partial_x^2 + \partial_y^2)}{2m^*} -  \mu \right)\tau_z + \alpha (i\partial_y\sigma_x-i\partial_x \sigma_y)\tau_z \nonumber \\
&+ E_Z \sigma_y + \Delta(x)\tau_+ + \Delta^*(x)\tau_-,
 \end{align}
where $\psi_{\uparrow,\downarrow}$ are the annihilation operators for electrons with spin up and down, $\boldsymbol{\sigma}$ and $\boldsymbol{\tau}$ are the Pauli matrices acting in the spin and particle-hole basis, respectively with $\tau_{\pm} = (\tau_x \pm i \tau_y)/2$. Here, $m^*$ is the effective electron mass, $\mu$ is the chemical potential, $\alpha$ is the spin-orbit coupling strength of InAs, $E_Z = g \mu_B B_{\parallel}/2$ is the Zeeman field strength due to the applied magnetic field along the junction ($y$ direction) and $\Delta$ is the proximity-induced pairing potential. The proximity-induced pairing potential is taken to be nonzero in the 2DEG below the superconducting leads and zero in the junction, i.e.,
\begin{equation}
\Delta(x) = \begin{cases}
\Delta e^{-i\varphi/2} & \text{for $ -(W_1/2+\Wso) < x < -W_1/2$},\\
0 & \text{for $ -W_1/2 < x < W_1/2$},\\
\Delta e^{i\varphi/2} & \text{for $ W_1/2 < x < W_1/2+\Wso$},
\end{cases}
\end{equation}
where $\varphi$ is the superconducting phase difference between the two superconductors, $W_1$ and $\Wso$ are the width of the junction and superconducting leads, respectively. The Zeeman field $E_Z$ is taken to be uniform throughout the system. Numerical simulations in this paper are done using experimental parameters: $m^* = 0.026m_e$~\cite{vurgaftman2001}, $\alpha = 0.1$ eV\AA~\cite{Ofarrell2018}, $\Delta = 0.15$ meV, $W_1 = 80$ nm and $\Wso = 160$ nm.

\textbf{Phase diagram.} We study the phase diagram of the system as a function of the in-plane Zeeman field, phase difference across the junction, and chemical potential.
In the limit of very wide superconducting leads ($\Wso \gg \xis$, where $\xis=\hbar v_{\rm F}/\pi \Delta$ is the superconducting coherence length and $v_{\mathrm F}$ is the Fermi velocity in the semiconductor), the topological phase transition depends on both the phase bias across the junction and the in-plane Zeeman field, with very weak dependence on the chemical potential~\cite{Pientka2017,Hell2017}. However, when the superconducting leads are narrow ($\Wso \ll \xis$), we expect the phase diagram to have a stronger dependence on the chemical potential and a weaker dependence on the phase bias, due to strong normal reflections from the superconductor edges.

\begin{figure*}
\includegraphics[width=\textwidth]{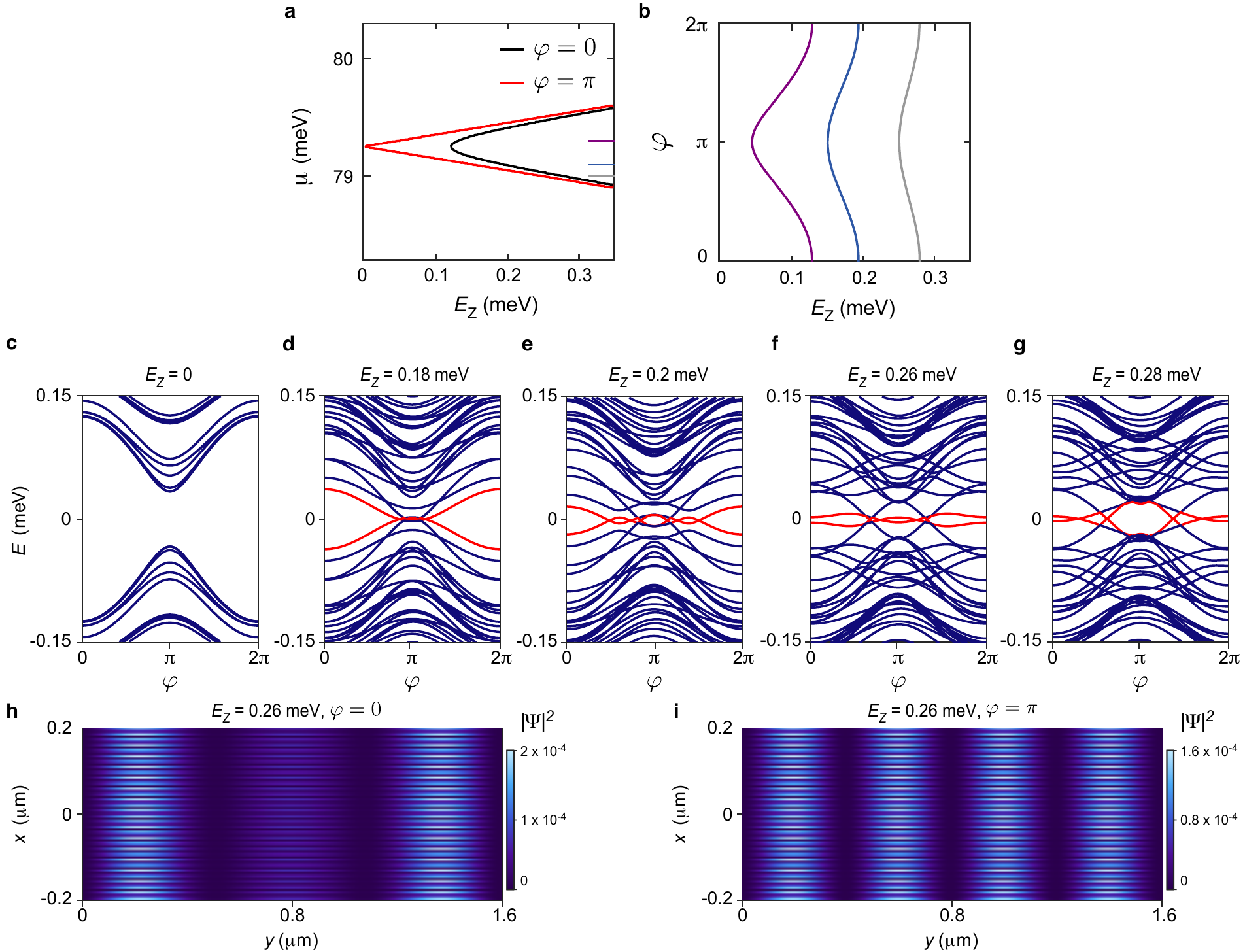}
\caption{\textbf{\boldmath Calculated topological phase diagrams and energy spectra.} \textbf{a}, Topological phase diagram as a function of the Zeeman energy $E_{\rm Z}$ and the 2DEG chemical potential $\mu$ for phase bias $\varphi=0, \pi$, calculated from the tight-binding Hamiltonian for JJ1 with infinite length (see Methods). The curves indicate the critical value of $E_{\rm Z}$ above which the system is tuned into the topological phase. \textbf{b}, Topological phase diagram as a function of $E_{\rm Z}$ and $\varphi$ for different values of $\mu$, as indicated by the horizontal ticks in panel \textbf{a}. The diagrams were calculated for a junction with width $W_1=80$ nm, superconducting lead width $\Wso=160$ nm, induced gap $\Delta=150~\mathrm{\mu}$eV and Rashba spin-orbit coupling constant $\alpha=100$ meV \AA. The length of the junction $L_1$ was assumed to be infinite in order to obtain a well-defined topological invariant, as described in the Methods. \textbf{c-g}, Calculated energy spectra as a function of $\varphi$ for different values of the Zeeman energy. The spectra were obtained for the same parameters used in panels \textbf{a} and \textbf{b}, except for $L_1=1.6~\mathrm{\mu}$m. The chemical potential $\mu$ was set to 79.1 meV (corresponding to the blue curve in panel \textbf{b}). For the chosen parameters, the system undergoes a topological transition at $\Ez=0.153$ meV for $\varphi=\pi$ and at $\Ez=0.195$ meV for $\varphi=0$. The lowest energy subgap states are shown in red and indicate two Majorana zero modes at the edges of the junction in the topological regime. As a function of $\Ez$ these states first reach zero energy at $\varphi=\pi$ and progressively extend in phase. At high values of $\Ez$ the Majorana modes oscillate around zero energy due to the finite size of our system which causes the Majorana wave functions to hybridize. This is particularly evident at $\varphi=\pi$, where the induced gap is minimized and the coherence length is maximized. \textbf{h, i}, Probability density $|\Psi|^2$ of the Majorana wavefunction calculated as a function of the spatial directions $x$ and $y$ in JJ1 for $\Ez=0.26$ meV and $\varphi=0,\, \pi$. $x$ is the coordinate in the width direction including the superconducting leads ($W_1+2\Wso=0.4~\mathrm{\mu}$m, with $x=0$ indicating the center of the junction), while $y$ is the coordinate along the length of the junction. The Majorana wavefunctions are localized in the $y$ direction at the edges of the junction when the lowest energy states in the spectrum are close to zero energy. In the $x$ direction the Majorana modes are delocalized below the superconducting leads, due to our geometry having $\Wso \ll \xis$.}
\label{figEDF1}
\end{figure*}

\begin{figure*}
\includegraphics[width=\textwidth]{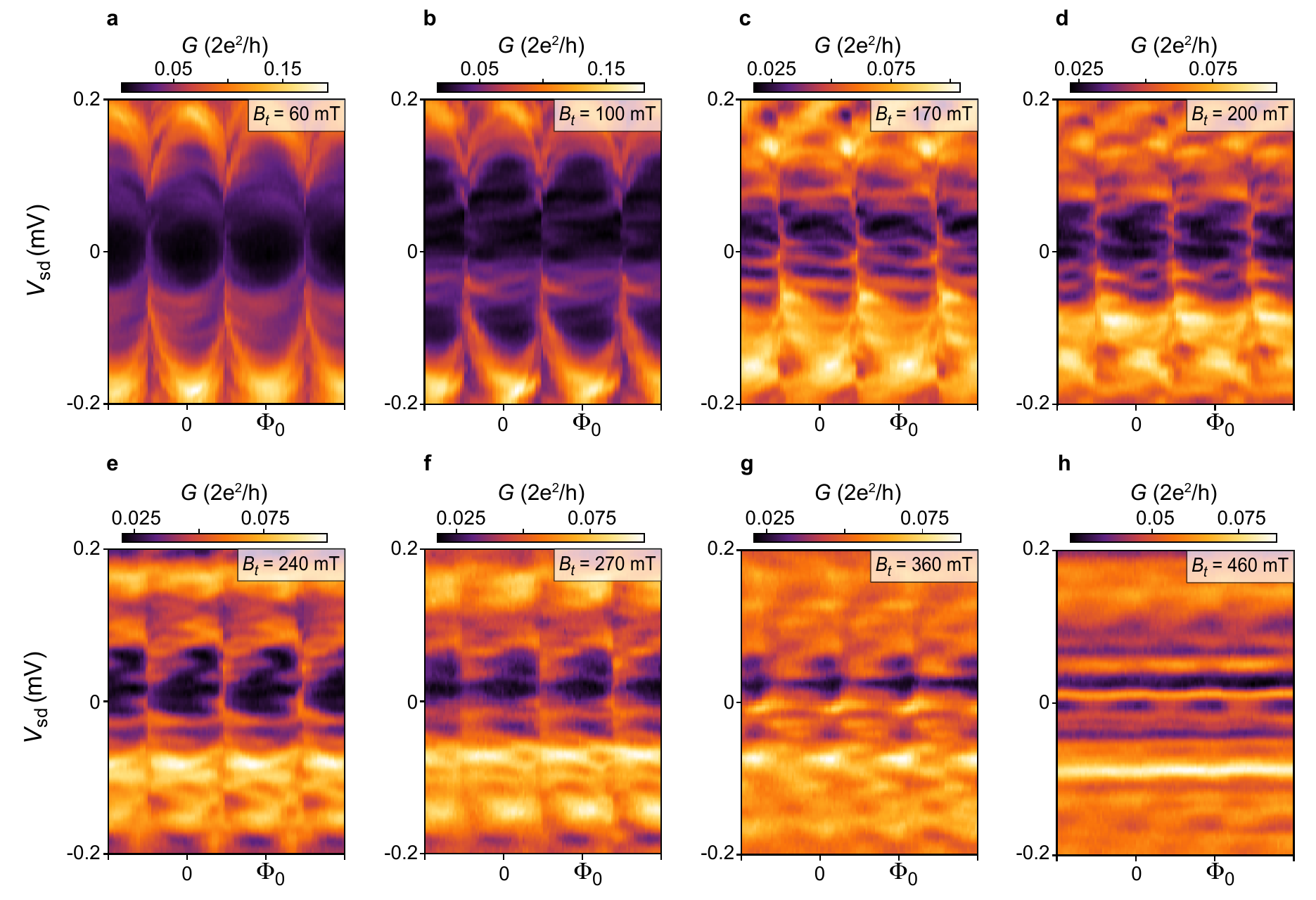}
\caption{\textbf{Transport spectroscopy in transverse field for device 1.} \textbf{a-h}, Differential conductance $G$ as a function of the magnetic flux, $\Phi$, threading the SQUID loop and source-drain bias, $\Vsd$, measured at different values of the transverse magnetic field $\Bt$ (applied in plane orthogonally to the junction) in device 1. Several ABSs enter the gap without sticking to zero energy. The induced gap collapses at $\sim 360$ mT.}
\label{figEDF2}
\end{figure*}

\begin{figure*}
\includegraphics[width=2\columnwidth]{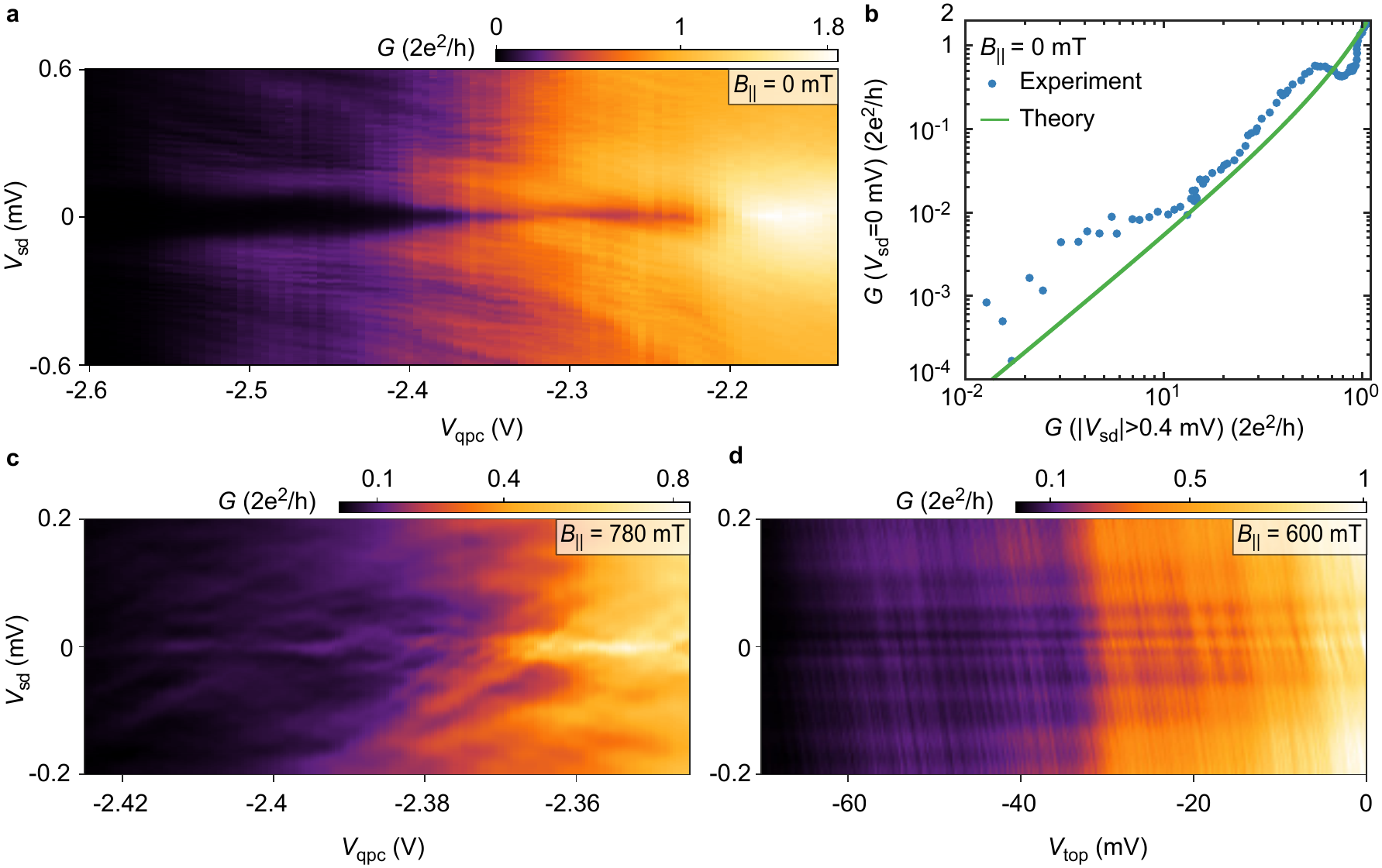}
\caption{\textbf{Quantum point contact characterization and stability of the zero-bias peak.} \textbf{a}, Differential conductance $G$ as a function of source-drain bias, $\Vsd$, and QPC voltage, $\Vqpc$, at zero field in device 1. \textbf{b}, Differential conductance at zero source-drain bias, $G(\Vsd = 0~\mathrm{mV})$, versus averaged differential conductance at finite source-drain bias, $G(|\Vsd|>0.4~\mathrm{mV})$. The green line is the theoretically predicted conductance in an Andreev enhanced QPC, $G_{\rm S}=2G_0 \frac{G_{\rm N}^2}{(2G_0-G_{\rm N})^2}$ (Ref.~\cite{Beenakker1992}), where $G_{\rm S}$ is the sub-gap conductance, $G_{\rm N}$ is the above-gap conductance and $G_0=2e^2/h$ is the quantum of conductance. No fitting parameters have been used. \textbf{c}, $G$ as a function of $\Vsd$ and $\Vqpc$ at parallel field $\Bp=780$ mT and phase bias $\varphi \sim 0.8 \pi$ for gate voltages $V_1=-110$ mV and $\Vt=-35$ mV. \textbf{d}, $G$ as a function of $\Vsd$ and $\Vt$ at $\Bp=600$ mT and $\varphi \sim 0$ for $V_1=-118.5$ mV and $\Vqpc=-2.366$ mV. In both panels \textbf{c} and \textbf{d}, the ZBP is robust against variation of the above gap conductance of about one order of magnitude.}
\label{figEDF3}
\end{figure*}

To obtain the phase diagram of the system, we perform numerical simulations by using the tight-binding version of the Hamiltonian [Eq.~\eqref{eq:HSM}]. We calculate the $\mathbb{Z}_2$ topological invariant ($Q = \mathrm{sign} [\mathrm{Pf}(H_{k_y=\pi}\tau_x)/\mathrm{Pf}(H_{k_y=0}\tau_x)])$~\cite{Tewari2012} for an infinitely long junction. The calculated phase diagrams are shown in Extended Data Fig.~\ref{figEDF1}a,b. The topological phase transition boundary, which separates the trivial region ($Q = 1$) at low Zeeman field from the topological region ($Q = -1$) at high Zeeman field, is marked by a gap closing at $k_y$ = 0. This topological transition corresponds to the transition between even (trivial) and odd (topological) number of subbands crossed by the Fermi level. For our case of narrow superconducting leads ($\Wso \ll \xis$), the $\mathbb{Z}_2$ topological phase diagrams are weakly dependent on the superconducting phase difference (see Extended Data Fig.~\ref{figEDF1}a). 
Extended Data Fig.~\ref{figEDF1}a also shows that for junctions with narrow superconducting leads, the dependence of the topological phase diagram on the superconducting phase difference is stronger for the case where the chemical potential is in the regime where the topological transition happens at a smaller Zeeman field. The phase diagram as a function of superconducting phase difference and chemical potential is shown in Extended Data Fig.~\ref{figEDF1}b.

\textbf{Energy spectra and Majorana wave function.} We calculate the energy spectra and Majorana wave functions by diagonalizing the tight-binding Hamiltonian for a finite-length system ($L_1 = 1.6$ $\mu$m). Extended Data Figures~\ref{figEDF1}c-g show the energy spectrum of the system as a function of the superconducting phase difference for several values of Zeeman field strengths. The spectrum shows a modulation with respect to the superconducting phase difference where the bulk gap assumes its minimum value at $\varphi = \pi$. In the limit where the junction is infinitely long, the gap closes at $\varphi = \pi$. This can be understood as follows. For the case of narrow superconducting leads ($\Wso \ll \xis$), electrons have to undergo multiple normal reflections from the edges of the superconductors before they can be Andreev reflected as it takes a length $\approx \xis$ for electrons to feel the presence of a gap. As a result, electrons feel a gap which is the weighted average of the left
and right superconducting gaps, i.e.,
\begin{align}
\tilde{\Delta} &= \frac{1}{2\Wso+W_1}\int_{-(W_1/2+\Wso)}^{W_1/2+\Wso} \Delta(x), \nonumber\\
&=  \Delta (e^{-i\varphi/2} + e^{i\varphi/2}) \frac{\Wso}{2\Wso+W_1},\nonumber\\
& = \Delta \frac{2 \Wso}{ 2 \Wso + W_1}\cos (\varphi/2), 
\end{align}
where the term $2 \Wso/ (2 \Wso+W_1)$ is the ratio of the superconductors width to the total width of the system. For a finite-length junction, the gap still has its minimum at $\varphi =\pi$ but does not close at zero Zeeman field as shown in Extended Data Fig.~\ref{figEDF1}c. The gap decreases with increasing Zeeman field strength and closes at $\varphi = \pi$ when the Zeeman field strength becomes sufficiently large. Furthermore, when the Zeeman field strength exceeds the critical value at which the topological phase transition happens, Majorana zero modes will appear at the end of the junction.  The Majorana zero modes first appear at $\varphi = \pi$ and as the Zeeman field strength is increased further, they extend in phase and are present for all values of superconducting phase difference. Extended Data Figs.~\ref{figEDF1}h,i show the probability densities of the lowest energy wave functions (corresponding to Majorana modes) calculated using the experimental parameters of our system. As can be seen in the figure, for our system with a length of $L_1 = 1.6$ $\mu$m, in some parameter regimes the Majoranas are localized at the system end and are well separated from each other.

\bibliographystyle{naturemag}
\bibliography{Bibliography}

\begin{figure*}
\includegraphics[width=1.2\columnwidth]{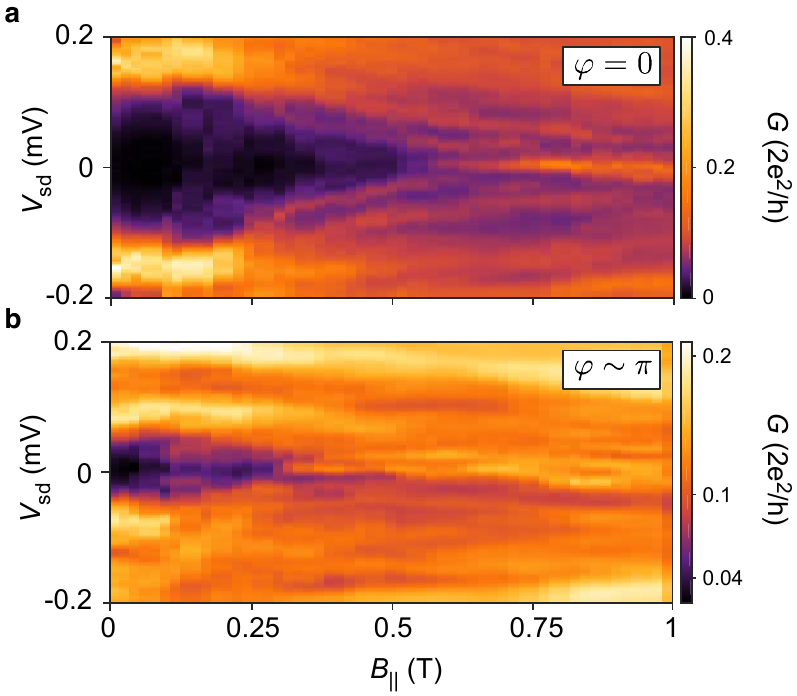}
\caption{\textbf{Field dependence of the zero-energy state for device 1.} \textbf{a-d}, Differential conductance $G$ as a function of source-drain bias, $\Vsd$, and parallel magnetic field, $\Bp$, for different values of phase bias $\varphi$ in device 1. The plots have been reconstructed from measurements similar to the ones shown in Fig.~\ref{fig2} of the Main Text. For $\varphi \sim \pi$ a ZBP forms at $\Bp=0.35$ T, while for $\varphi=0$ it appears at $\Bp=575$ mT. The ZBP at $\varphi \sim\pi$ oscillates and moves away from zero energy as the field is increased.}
\label{figEDF4}
\end{figure*}

\begin{figure*}
\includegraphics[width=\textwidth]{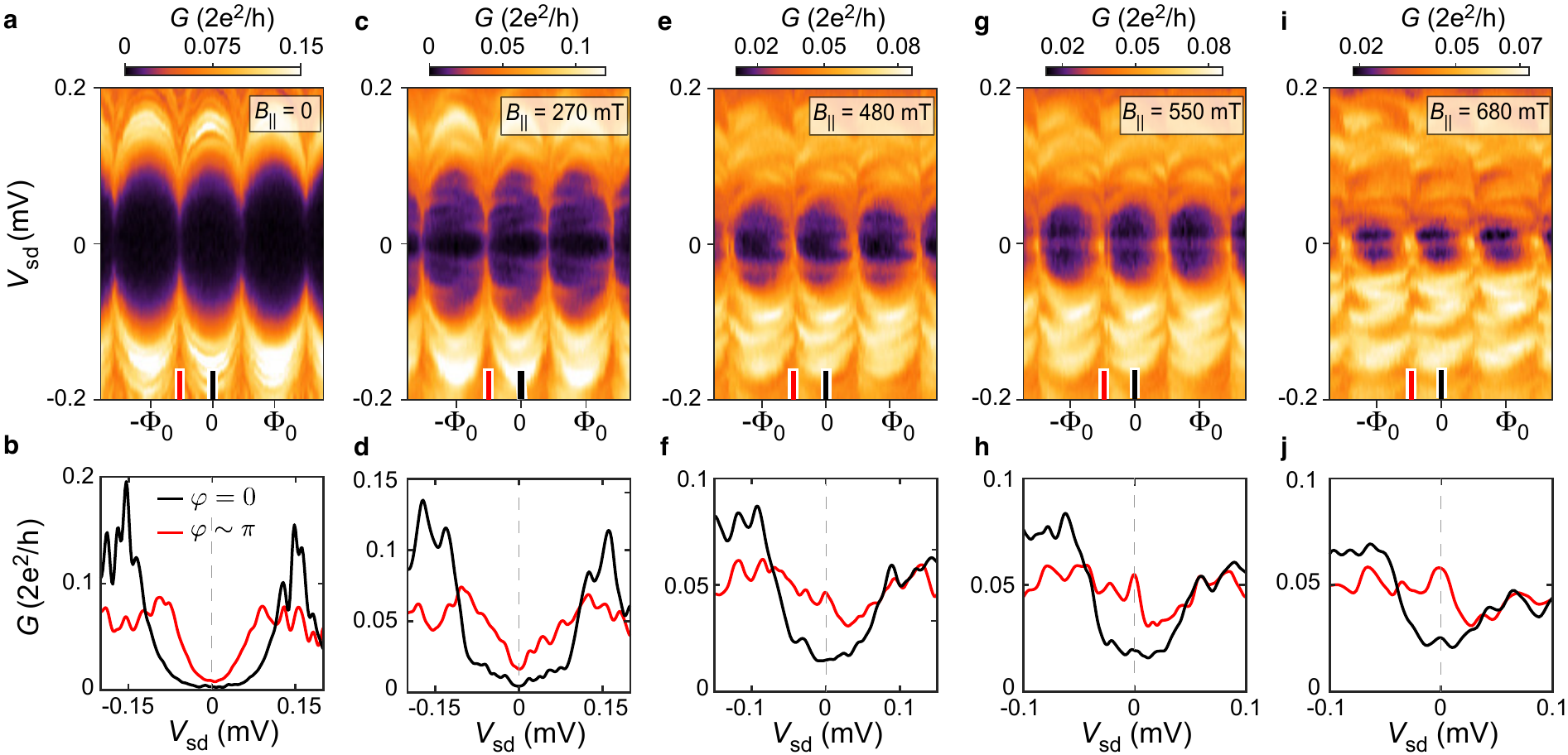}
\caption{\textbf{\boldmath Tunneling spectroscopy at lower tunneling transmission.} \textbf{a, c, e, g, i},  Differential conductance $G$ as a function of magnetic flux, $\Phi$, threading the SQUID loop and source-drain bias, $\Vsd$, measured at different values of parallel magnetic field $\Bp$ in device 1 ($W_1=80$ nm). The QPC was tuned to reduce the above-gap conductance by a factor $\sim 3$ with respect to the one measured in the regime presented in the Main Text. At zero field the sub-gap conductance is suppressed. Colors extrema have been saturated. \textbf{b, d, f, h, j}, Conductance line cuts vs. $\Vsd$ taken at $\varphi = 0,\pi$, as indicated by red and black ticks in the top panels. The grey dashed lines indicate $\Vsd=0$.}
\label{figEDF5}
\end{figure*}

\begin{figure*}
\includegraphics[width=0.8\textwidth]{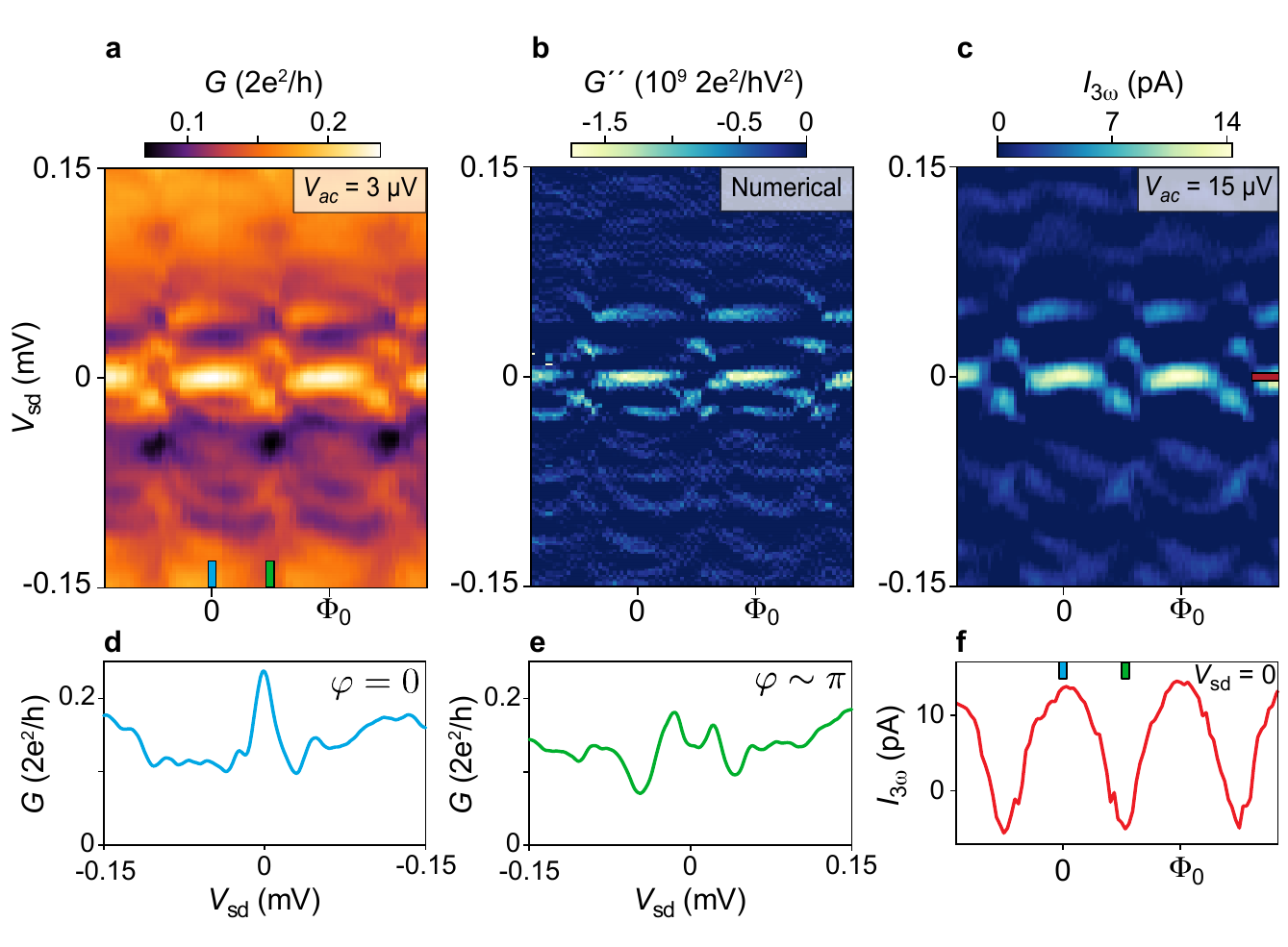}
\caption{\textbf{Measurement of the third harmonic of the current.} \textbf{a}, Differential conductance $G$ as a function of source-drain bias $\Vsd$ and magnetic flux $\Phi$ at parallel magnetic field $\Bp=850$ mT measured with an excitation amplitude $V_{\rm ac} = 3 ~\mathrm{\mu}$V in device 1. \textbf{b}, Numerical second derivative of the conductance $\dG=(\partial^2 G/\partial \Vsd^2)|_{\Vsd}$ as a function of $\Vsd$ and $\Phi$ calculated from the data shown in \textbf{a}. \textbf{c}, Third harmonic of the current $\It$ versus $\Vsd$ and $\Phi$ measured by the lock-in amplifier using an excitation $V_{\rm ac}=15 ~\mathrm{\mu}$V, as explained in the Methods. In order to increase the signal-to-noise ratio, the amplitude of the excitation has been chosen to be greater than the temperature-limited full width at half maximum of a Lorentzian feature, i.e., $V_{\rm ac}\gtrsim 3.5\, k_{\rm B}T$, where $k_{\rm B}$ is the Boltzmann constant and $T\sim 40$ mK is the electron temperature in our device. Most of the features present in panel \textbf{b} are reproduced in panel \textbf{c}. \textbf{d, e}, Line cuts of $G$ as a function of $\Vsd$ taken at $\varphi=0,\,\pi$ as indicated by the ticks in panel \textbf{a}. \textbf{f}, $\Itoo$ as a function of $\Phi$: a positive value of $\Itoo$ indicates a ZBP in $G$. See Methods for further details.}
\label{figEDF6}
\end{figure*}

\begin{figure*}
\includegraphics[width=\textwidth]{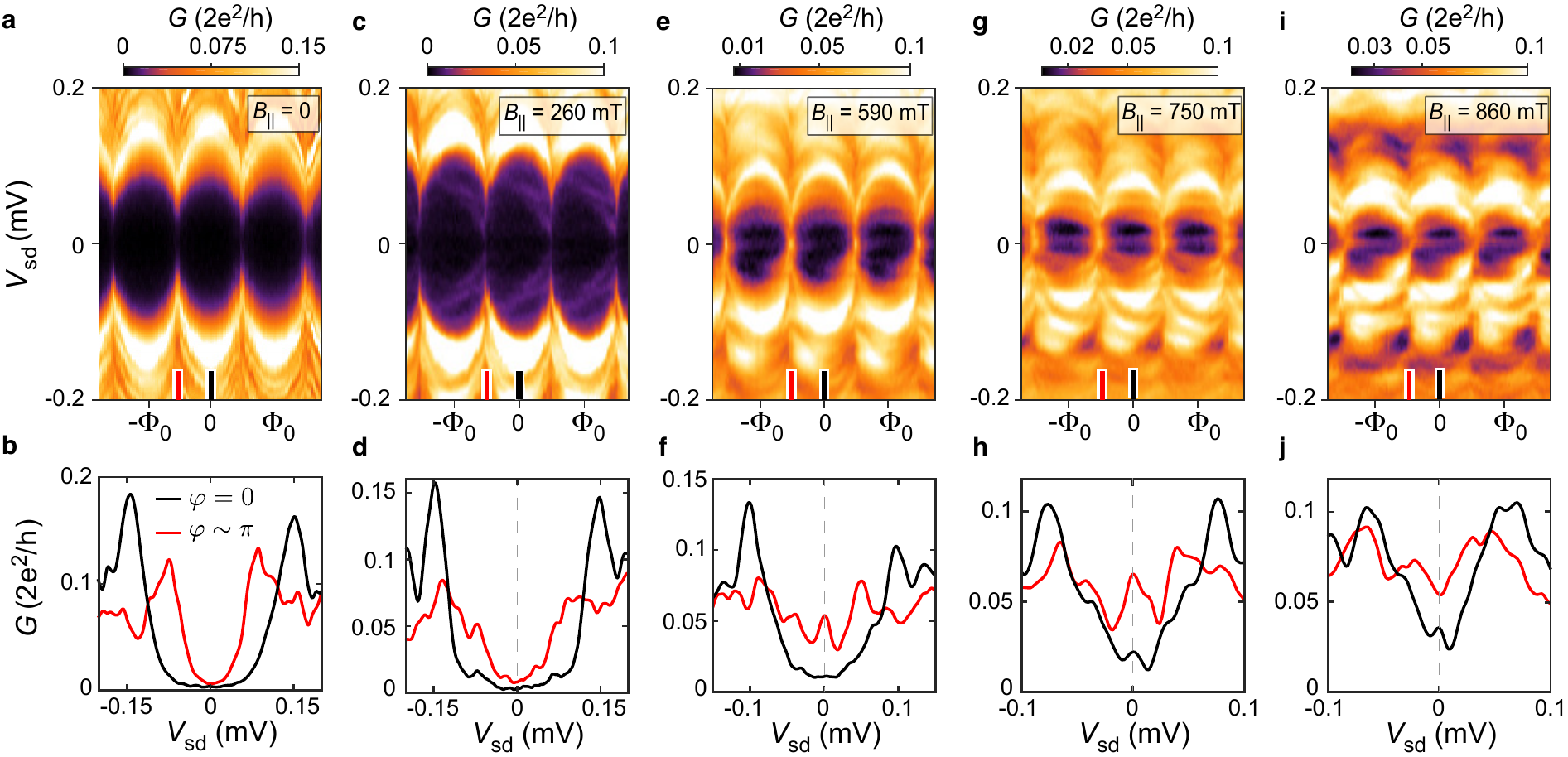}
\caption{\textbf{\boldmath Tunneling spectroscopy in device 2 ($W_1=80$ nm).} \textbf{a, c, e, g, i},  Differential conductance $G$ as a function of magnetic flux, $\Phi$, threading the SQUID loop and source-drain bias, $\Vsd$, measured at different values of parallel magnetic field $\Bp$ in device 2 ($W_1=80$ nm). Colors extrema have been saturated. \textbf{b, d, f, h, j}, Conductance line cuts vs. $\Vsd$ taken at phase bias $\varphi = 0,\pi$, as indicated by red and black ticks in the top panels. The grey dashed lines indicate $\Vsd=0$.}
\label{figEDF7}
\end{figure*}

\begin{figure*}
\includegraphics[width=0.8\textwidth]{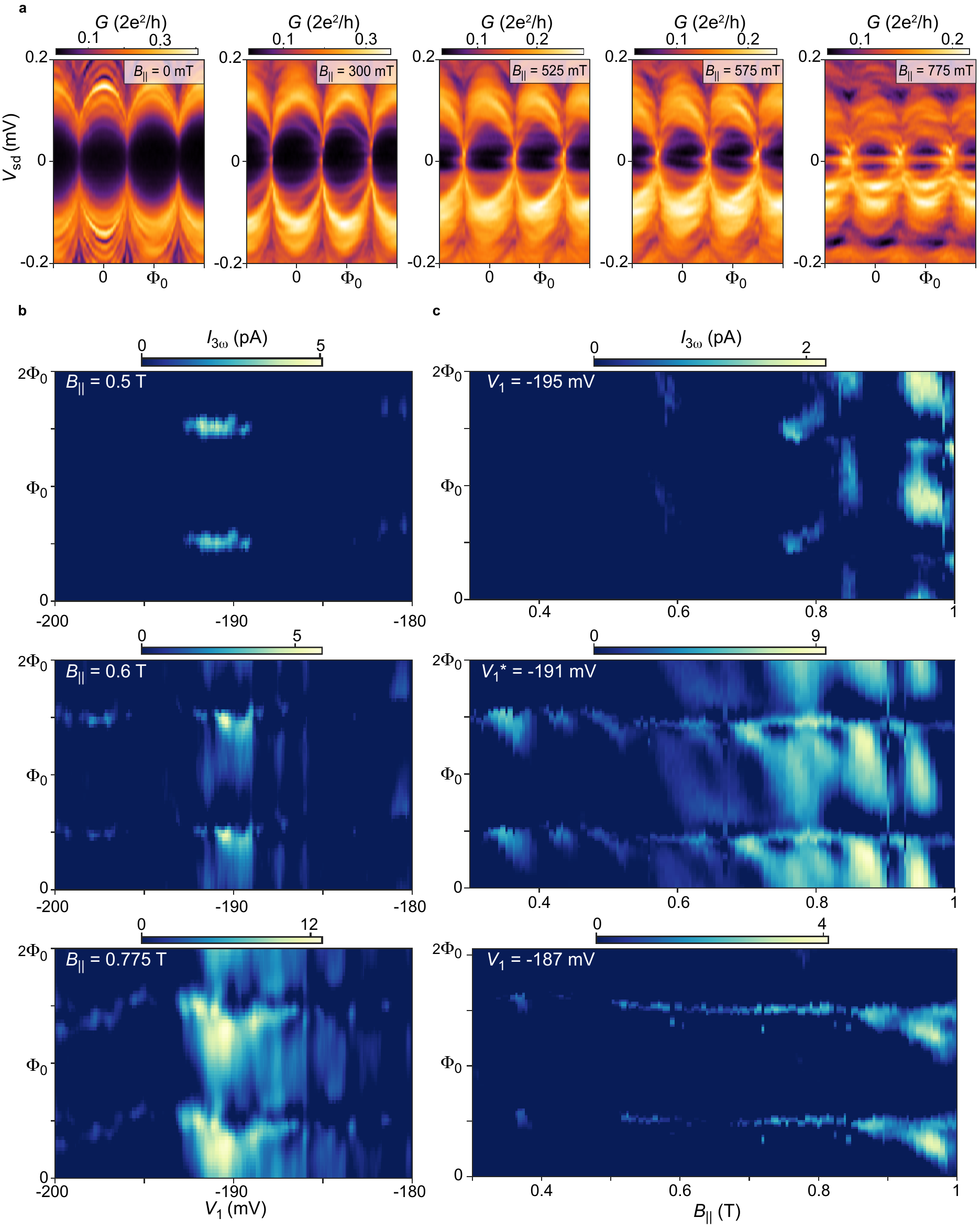}
\caption{\textbf{\boldmath Zero-bias peak stability in device 2 ($W_1=80$ nm).} \textbf{a},  Differential conductance $G$ as a function of magnetic flux, $\Phi$, piercing the SQUID loop and source-drain bias, $\Vsd$, measured at different values of parallel magnetic field $\Bp$ in device 2 at top gate voltage $V_1=-191$ mV. \textbf{b}, Third harmonic of the current $\Itoo$ measured by the lock-in amplifier at zero bias as a function of $V_1$ and $\Phi$ for different values of $\Bp$. $\Itoo \propto -\dGo=-(\partial^2 G/\partial V^2)|_{\Vsd=0}$, as shown in the Methods and Extended Data Fig.~\ref{figEDF6}. A positive value of $\Itoo$ corresponds to a ZBP in conductance as a function of $\Vsd$. As $\Bp$ is increased, the ZBP expands in phase and in $V_1$ range, consistent to what is observed in Fig.~\ref{fig3} of the Main Text for device 1. \textbf{c}, Third harmonic of the current $\Itoo$ measured by the lock-in amplifier at zero bias as a function of $\Bp$ and $\Phi$ for different values of $V_1$. A positive value of $\Itoo$ corresponds to a ZBP in conductance as a function of $\Vsd$. The critical field at which the ZBP first appear is minimized at phase bias $\varphi=\pi$. The behavior of the ZBP is tuned by $V_1$ and is qualitatively consistent with the topological phase diagrams shown in Extended Data Figs.~\ref{figEDF1}a,b.}
\label{figEDF8}
\end{figure*}



\begin{figure*}
\includegraphics[width=\textwidth]{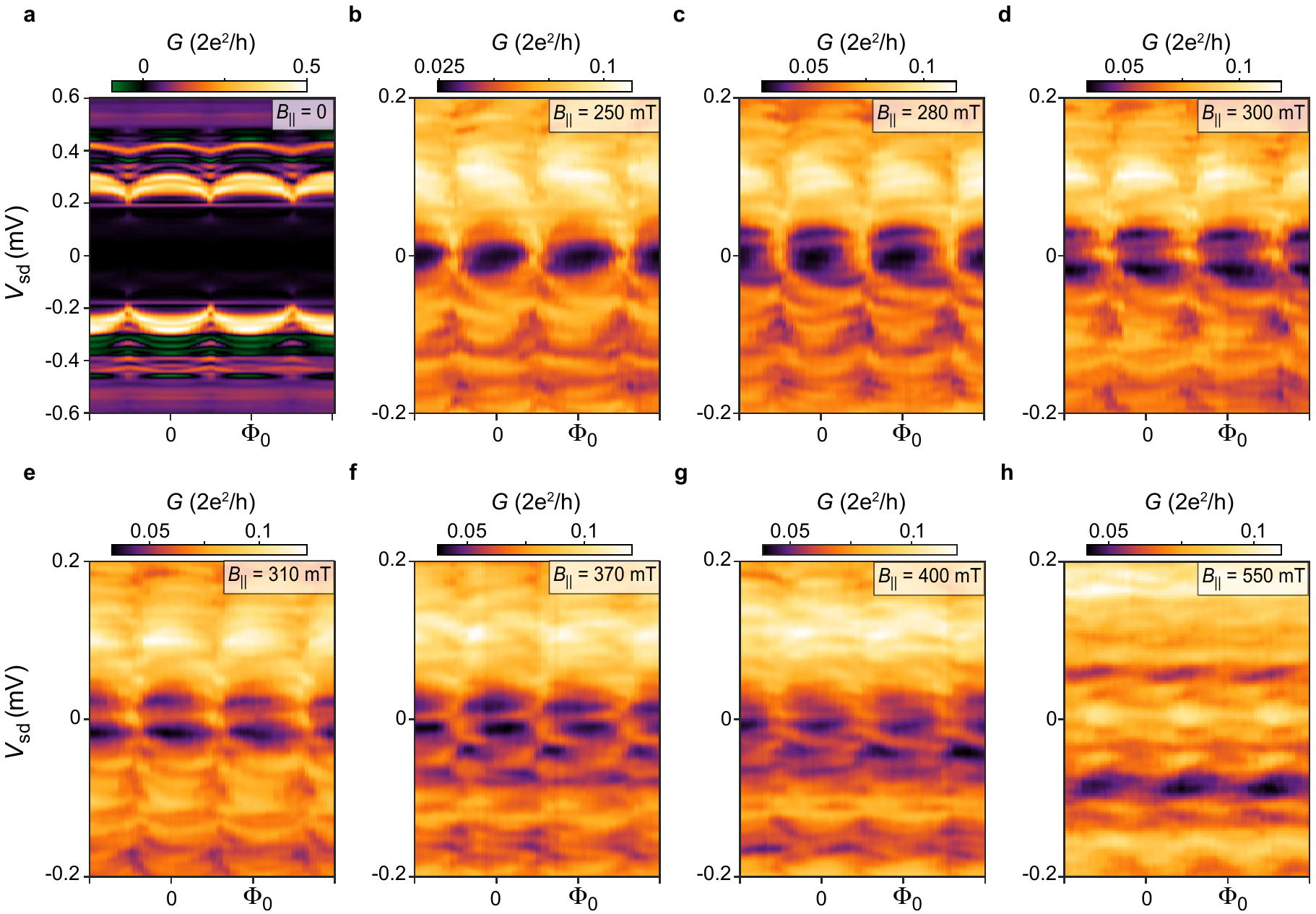}
\caption{\textbf{\boldmath Tunneling spectroscopy in device 3 ($W_1=120$ nm).} \textbf{a-h},  Differential conductance $G$ as a function of magnetic flux $\Phi$ and source-drain bias $\Vsd$ measured at different values of parallel magnetic field $\Bp$ in device 3 ($W_1=120$ nm). In Device 3 spectroscopy was performed with a QPC forming a tunnel barrier between the top edge of JJ1 and a wide planar Al lead, following the approach of Refs.~\citenum{Suominen2017,Nichele2017}. At $\Bp=0$ the superconducting probe generates a flux-independent gap $\Delta_{\rm probe}^{\star}\simeq 200~\mathrm{\mu}$eV added to the junction gap $\Delta\simeq 100\mathrm{\mu}$eV, together with characteristic features of negative differential conductance, as visible in panel \textbf{a}. When a moderate parallel field is applied, the superconducting gap below the superconducting plane softens, creating a finite density of states at zero energy. This feature allows the Al plane to be used as an effective normal lead that can probe discrete states close to zero energy in the junction. At $\Bp = 250$ mT, we can see a complete phase modulation of $\Delta$, indicating that the Al plane gap is already soft (see panel \textbf{b}). As the field is increased, two ABSs move towards zero energy, forming a ZBP first localized at phase bias $\varphi=\pi$ and then extending up to $\varphi = 0$, as shown in panels \textbf{c-f}. At higher fields, the induced gap collapses (panels \textbf{g,h}). The lower value of $\Delta$ and critical field compared to that observed in devices 1 and 2 are presumably due to the larger width of the junction.}
\label{figEDF9}
\end{figure*}

\begin{figure*}
\includegraphics[width=0.7\textwidth]{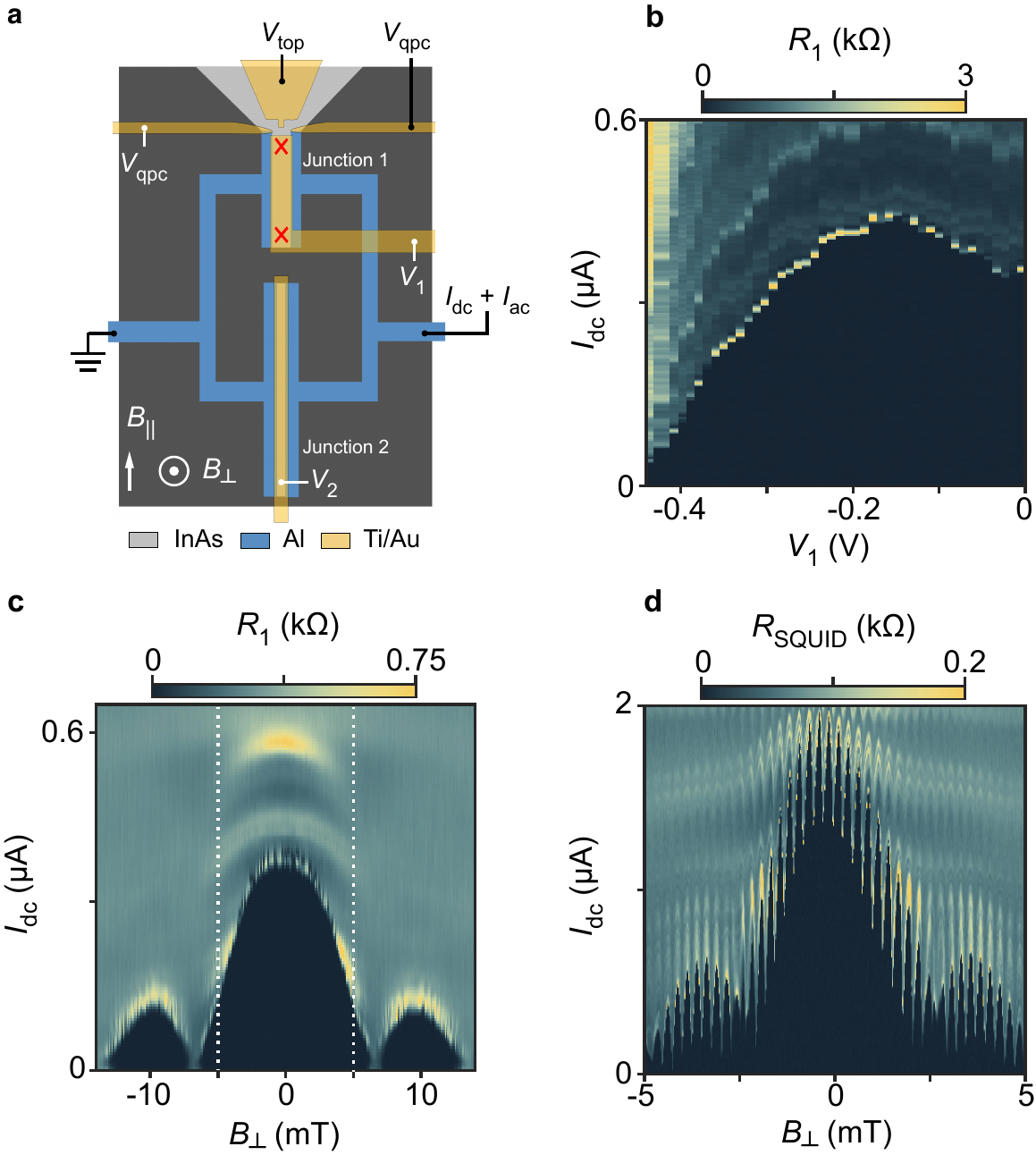}
\caption{\textbf{\boldmath Behavior of the Josephson critical current at $\Bp=0$.} \textbf{a}, In order to investigate the behavior of the Josephson current in our device, we measured the differential resistance $R=dV/dI$ of the SQUID with a conventional four-probe technique by applying an AC current bias $I_{\rm ac}<5$ nA, superimposed to a variable DC current bias $I_{\rm DC}$, to one of the superconducting leads of the interferometer. During these measurements the QPC was pinched off at $\Vqpc=-5$ V. The Josephson critical current of JJ1 can be measured independently by pinching off JJ2.\textbf{b}, Differential resistance $R_1$ of JJ1 as a function of the DC bias current $I_{\rm dc}$ and gate voltage $V_1$ measured in device 2. The region of zero resistance indicates that a dissipationless Josephson current due to Cooper pair transport is flowing through the junction. \textbf{b}, $R_1$ as a function of $I_{\rm dc}$ and the out-of-plane field $B_{\perp}$ displaying a characteristic Fraunhofer pattern, with a periodicity compatible with the area of JJ1 $W_1 \times L_1\simeq 0.13~\mathrm{\mu m^2}$. For both the measurements displayed in panels \textbf{a} and \textbf{c}, JJ2 was pinched off by setting the gate voltage $V_2=-1.5$ V. \textbf{d}, When JJ2 was open ($V_2=0$), the differential resistance of the SQUID $R_{\rm SQUID}$ showed periodic oscillations (periodicity of 250 $\mathrm{\mu}$T, consistent with the area of the superconducting loop, $\sim 8~\mathrm{\mu m^2}$) superimposed to the Fraunhofer patterns of both junctions. The ratio between the critical currents of the junctions at zero field is extracted from the average value of the SQUID critical current and the semi-amplitude of the SQUID oscillations, resulting in $I_{\rm c,2}(0)/I_{\rm c,1}(0)=5.2$.}
\label{figEDF10}
\end{figure*}

\begin{figure*}
\includegraphics[width=\textwidth]{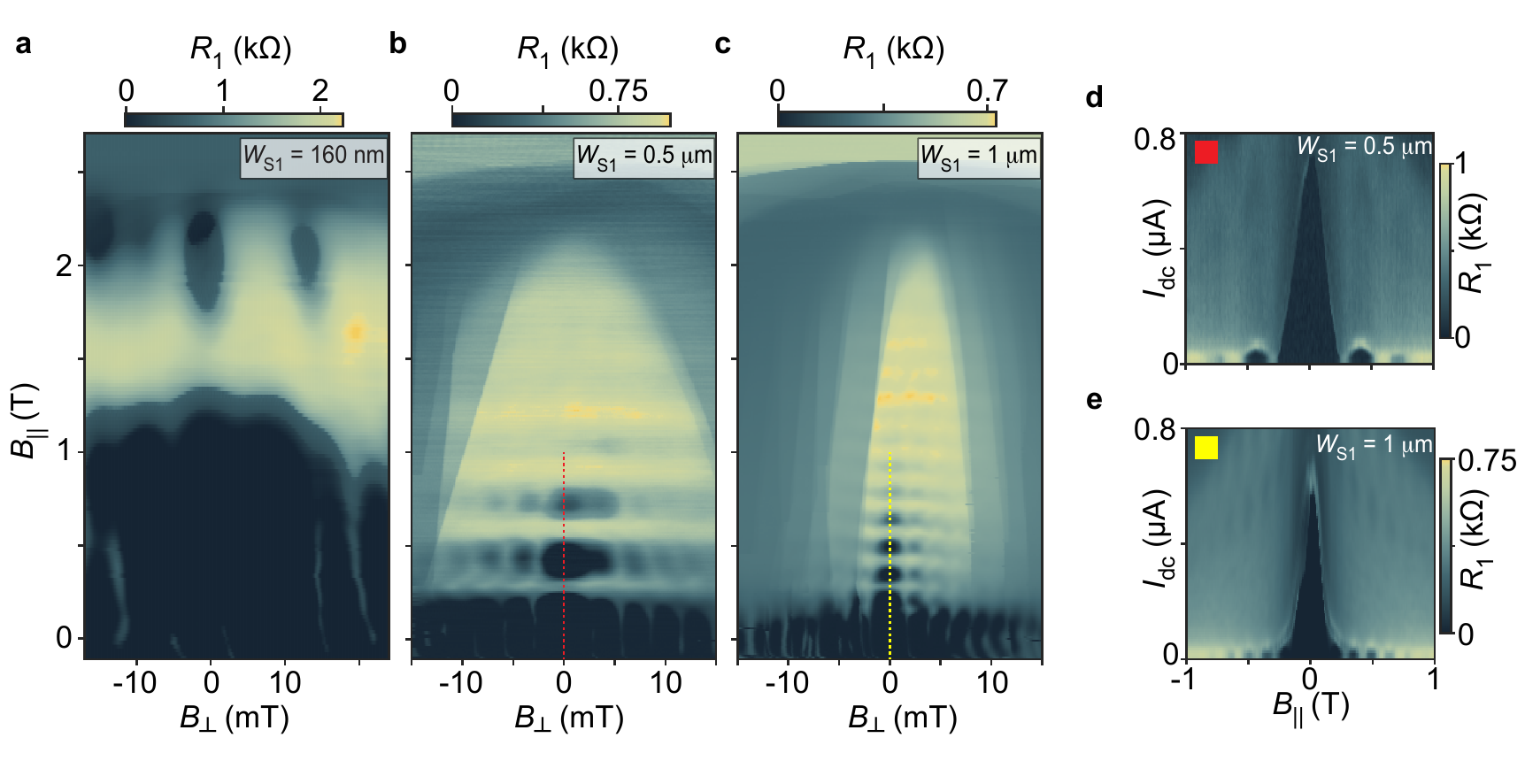}
\caption{\textbf{Trivial Josephson current revival in parallel field.} \textbf{a-c}, Differential resistance $R_1$ of JJ1 as a function of out-of-plane magnetic field $B_{\perp}$ and parallel magnetic field $\Bp$ measured in devices 2, 4 and 5 for DC current bias $I_{\rm dc}=0$ and AC current $I_{\rm ac}=5$ nA. All the devices are characterized by width $W_1=80$ nm and length $L_1=1.6~\mathrm{\mu}$m, while the width of the superconducting leads $\Wso$ (see Fig.~\ref{fig1}a of the Main Text) is varied. Device 2 is characterized by $\Wso=160$ nm, device 4 has $\Wso=500$ nm and device 5 $\Wso=1~\mathrm{\mu}$m. In the case of $\Wso=160$ nm (panel \textbf{a}), JJ1 becomes resistive at $\Bp \sim 1.1$ T and a supercurrent revival is observed above 2 T. The normal state transition of the epitaxial Al occurs at $\Bp \sim 2.4$ T, where the junction resistance reaches a value $\sim 1$ k$\Omega$ without any magnetic field dependence. When $\Wso$ is increased, the supercurrent revivals occur at lower values of $\Bp$ and show an evident periodicity of $\sim 300$ mT and $\sim 150$ mT for $\Wso=500$ nm (panel \textbf{b}) and $\Wso=1~\mathrm{\mu}$m (panel \textbf{c}), respectively. \textbf{d, e}, $R_1$ as a function of $I_{\rm dc}$ and $\Bp$ measured in devices 4 and 5 for $B_{\perp}=0$, as shown by the dashed lines in panels \textbf{b} and \textbf{c}. The Josephson current shows a clear Fraunhofer pattern due to orbital effects of the in-plane field penetrating the proximitized 2DEG below the Al leads~\cite{Pientka2017}. The measurements were performed with the QPC and junction 2 pinched off.}
\label{figEDF11}
\end{figure*}

\end{document}